# XDoser, a Benchmarking Tool for System Load Measurement using Denial of Service Features

AKM Bahalul Haque, Rabeya Sultana, Mohammad Sajid Fahad , MD Nasif Latif and Md. Amdadul Bari

Department of Electrical and Computer Engineering, North South University, Dhaka, Bangladesh

*ABSTRACT*

*Technology has developed so fast that we feel both safe as well as unsafe in both ways. Systems used today are always prone to attack by malicious users. In most cases, services are hindered because these systems cannot handle the amount of over loads the attacker provides. So, proper service load measurement is necessary. The tool that is being described in this paper for developments is based on the Denial of Service methodologies. This tool, XDoser will put a synthetic load on the servers for testing purpose. The HTTP Flood method is used which includes an HTTP POST method as it forces the website to gather the maximum resources possible in response to every single request. The tool developed in this paper will focus on overloading the backend with multiple requests. So, the tool can be implemented for servers new or old for synthetic test endurance testing.*

## 1. Introduction

Systems developed nowadays are being used by various types of user. Before releasing a system for usage, it needs to be tested properly. These testing provide the owners a confirmation about the reliability of the system. More and more users use the system the more the vulnerability increases. Moreover, the number of users also increases with time. All the users using the system does not have a very noble thing in mind. They might want to disrupt the system services. Denial of Service attack is one of the most effective ways to disrupt the services. That is why various benchmarking tool is used to check whether the website can handle a considerable amount of load.

The main aim of DOS attack is disruption of services by consuming the bandwidth of legitimate customer. This attack is done by sending a stream of illegitimate packets to certain victim to cut off the supply of the authentic packets. During the last few years, several prominent websites were the victims of such attacks [1]. A distributed denial of service (DDoS) is a type of attack that is performed from distributed system to disrupt the service. This method is becoming very complex and popular day by day. Many security tools have been proposed to fight the problem [2]. The major goals of DDoS attack are to consume bandwidth and overwork the server. Over the years, the most common choice has been the TCP SYN flood, with the ping flood a distant second. Application layer attacks are increasing, such as HTTP GET request floods and 'mail bombs' or floods from spam networks. DNS based attacks (attackers flood DNS servers with bogus but well-formed requests) are also quite popular. A normal request flood may overwork the server [3]. This research is mainly focused on improving server load capacity and endurance. The main purpose of our study is the understand DDoS attacks as a whole and eliminate possible server hacks. The tool developed in this paper will be able to create a synthetic load on servers to check the capacity it can handle.

Lower graphs showcase the DDOS attacks. The results of the first two quarters of the year 2018, the peak activity in Q2 2018 was observed in the middle of April 2018. The quarter's lowest number was observed respectively 24$^{th}$ May and June 17$^{th}$. During the Sunday and Tuesdays of the second quarter of 2018, the number was recorded at 14.99% and 17.49% respectively. According to the attacks logged the attack came down to 12.37% on Thursday. Overall, it can be observed from the graph, in the period of April-June the attack distribution over the days of the week was more even at the beginning of the year. The longest attack in Q2 lasted for 158 hours and for almost 11 days slightly shorter than the previous quarter's record of 297 hours that is 12.4 days. This time the focus of the preserving hackers was an IP address belonging to China Telecom [4] [5]. These are represented graphically below.

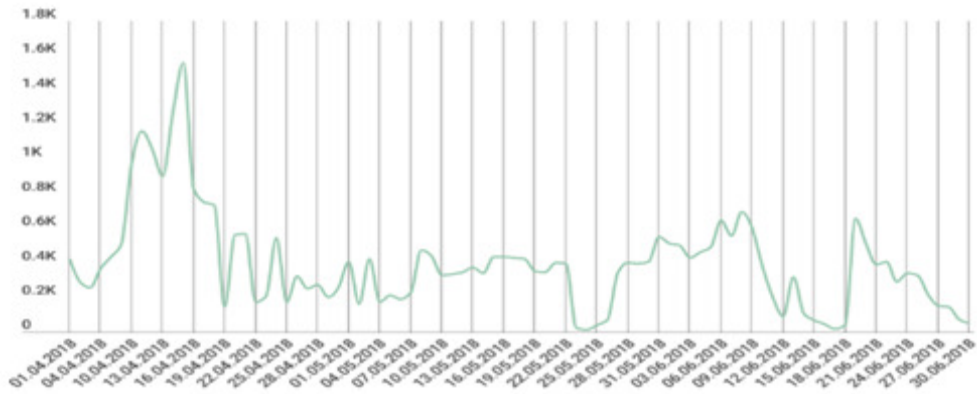

Figure 1: Dynamics of the number of DDoS attacks, Q2 2018

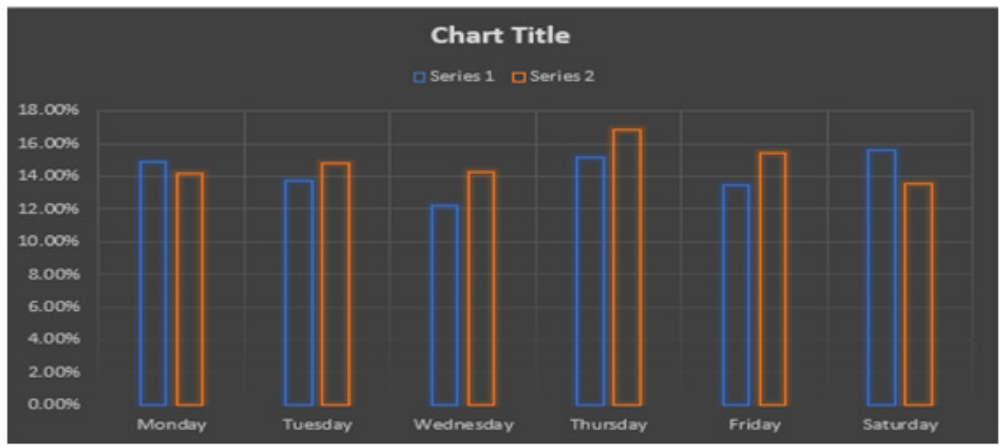

Figure 2: Distribution of DDOS attacks by day of the week, Q1 and Q2 2018

## 2. RELATED WORKS

As described in the previous section of this paper, for testing load capacity and service unavailability; it is important to discuss about DDoS in different aspects. It has always been one of the most effective ways to conduct unethical activities.

For this reason, building an effective tool which can perform better DDoS that currently available conventional tools and testing the system against the attack using the tool in a very effective way. So, the types and evolution of DDoS attack will be described in this section. If the evolution can be tracked, it can be one of the pillars for building the tool.

Several works related to DDoS have been done for years. The authors in [6] showed the risk of botnet-based attacks especially on the application layer of the system architecture and also the large-scale loss it can cause. Most DoS based bandwidth attacks are normally done with one single computer. However, trying to cause damage by repetitive usage can be detected easily. Even today, requests are more spread out over multiple types and sizes available. Particularly focusing on a large file can be faster, but it will be easily detected as an anomaly. The paper states about Distributed Reflector Attack. In such an attack the user hides the source of the attack, which is essential to the problem. It ends with solutions during that time and future developments. This helped us understand the attack in detail and the defense mechanism used years back and to counteract accordingly.

Another paper [7] suggests a different type of DDoS attack called the Silent Attack. The name is given as it poses as congestion to avoid detection as long as possible. The point was to be undetectable as long as possible. The use of TCP traffic one of our major foundation that the idea came from. Another paper [8] from 2003 was used to comprehend the evolution and the understanding of DDoS attack. Papers like this helped us to see the change in the timeline and the defenses used to better counteract with the solutions that might evolve to overcome the DOS tool. This paper focuses on making an efficient attack and improves on what was present.

The authors of the paper [9] used the idea of amplification of the strength of the DDoS attack. Unprotected computers are used as a bot against the target server. The attack amplification factor has other variables to work with. Understanding this paper made it easier to theorize and later apply the work.

## 3. MATERIALS AND METHODOLOGY

### 3.1 MATERIALS

For making this tool, the following tools were used:

- Protocols: HTTP
- Frameworks: Spring Boot
- Database: MySQL using XAMPP
- Server: Apache

The full form of HTTP is the Hypertext Transfer Protocol. It is a stateless protocol. It has request methods like GET, HEAD, POST, PUT and DELETE. It has the capability of dealing with data representation that allows any website to be built independently during transmission of data. The POST method is used to send the request to the Apache server [10].

"Spring Boot" framework is used to implement the backend. It automatically configures Spring as well as third-party libraries whenever possible. Moreover, it does not require for XML configuration.

MySQL has a distinct storage-engine framework that allows for flawless high performance. It can handle millions of queries . XAMPP contains the Apache module. In this paper, Apache is used as the server for the following three reasons[11][12].

 Firstly, Apache is the most popular Web server which is running at a faster rate in today's generation [13].

Secondly, Apache is a fully featured and high-performance Web server which means that its functionality, efficiency, and speed is better than any other server.

Thirdly, the source code of Apache is available which means that anyone can use it and it enables us to make changes to the code to improve its performance [14]. IntelliJ IDEA is an integrated development environment (ide). It indexes the whole project. It analyses everything and creates the syntax tree. In the code, wherever you put the cursor, it identifies the specific codes and what necessary changes can be done here [15].

### 3.2 METHODS

There are many types of DDoS attack like UDP Flood, SYN Flood, HTTP Flood and so on. The attack type used in this paper is the HTTP Flood method [15]. A GET request is used to retrieve standard and static content like images. It means that if GET request is used to hit the front end, then it cannot create a heavy load on the front end. So, the attack would not be an efficient one. Moreover, it cannot dynamically generate resources. HTTP GET request works on the front end.

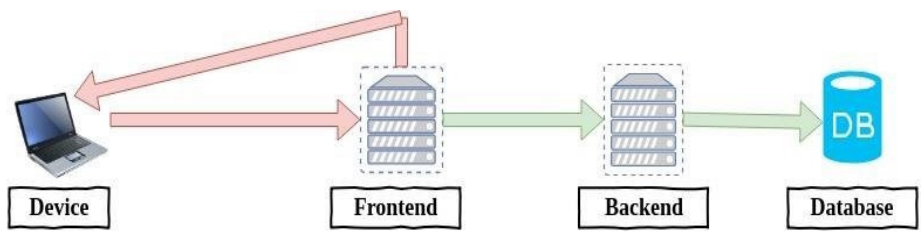

Figure 3. Typical dos Attack using GET request on the front end

The HTTP POST method is used to send the DOS attack on the backend. It means HTTP GET request are sent continuously in the form of searching the data. As the data is searched frequently, then this will create a load on the backend. Ultimately, it will create the load on the website. A generalized picture regarding the DOS attack on backend will look like the figure below:

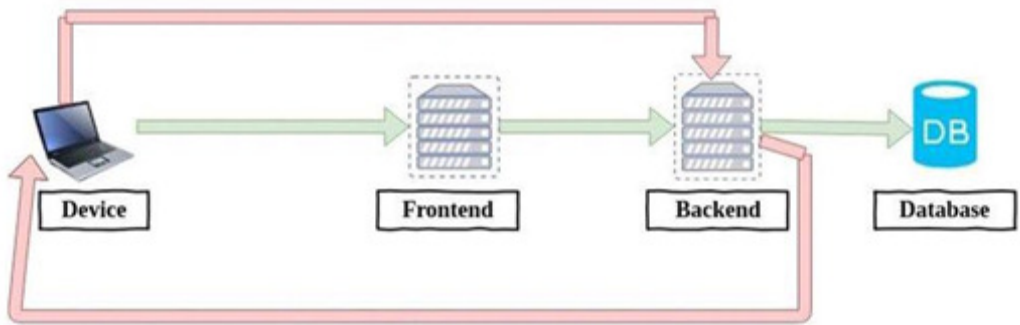

Figure 4. General Architecture of XDoser

The name of the website is Webdoser. The table in the database was named the course model. The table contains details like username(course name), address, age, full name, and institute name. Webdoser contains a folder named "Course Dao" that will allow the user to search for the course name using HTTP request. Now, for the search operation to take place, an object needs to be created for calling any HTTP request like search, delete and so on. A single object (client) was created using OkHttpClient so that the user can search for the course name in the database [16]. The search page created is given on the next page:

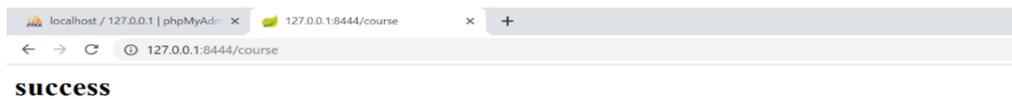

Figure 5. Input given by the user

Figure 6. HTML page for a search match

The word "SUCCESS" has been displayed on the screen as the searched data has been matched. Now, XDoser will create a severe load on the backend due to the repetitive search requests.

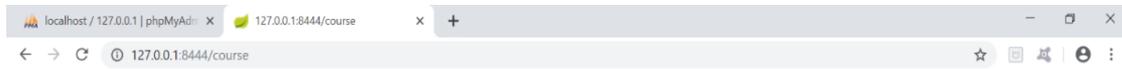

Figure 7. Using a different course name

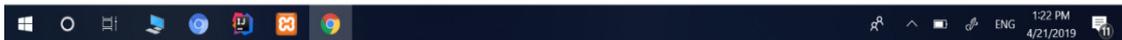

Figure 8. HTML page for failed request

"Failed" has been displayed. It means that this particular data is not in the database. Even though XDoser will create a load on the backend because the user is continuously sending HTTP POST request for searching a wrong data. Since repetitive searching is going on, so XDoser is creating load concurrently on the backend.

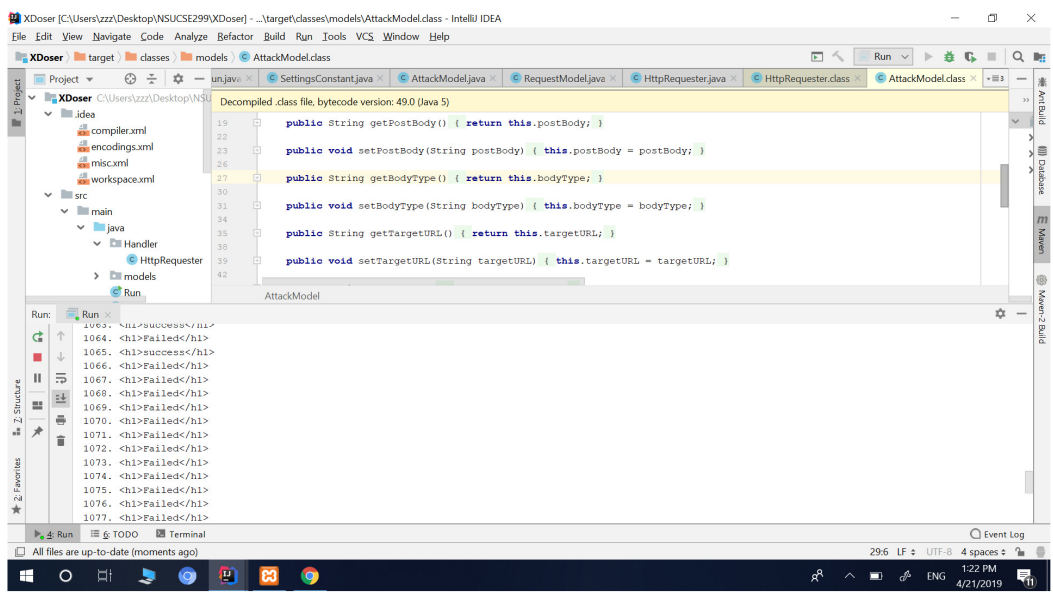

**Figure 9**. XDoser attack showing success and failed results proving that load has been created as the failed ratio is much higher

If the search result matches or not, it will show "Failed" due to the effect of XDoser as shown in Figure 10. Thus, the user will focus on overloading the backend with multiple requests that are each as processing-intensive as possible to check the maximum load it can bear. That is why HTTP flood [7] method using POST requests tend to be the most resource-effective and flawless method from the user's view. The thread will run concurrently and ultimately the server will go down. The Thread code is given below:

```
public static void main(String[] args) {
    final HttpRequester httpRequester = new HttpRequester();
    try {
        for (int i = 0; i < 50; i++) {
            Thread thread = new Thread(new Runnable() {
                public void run() {
                    try {
                        attack();
                    } catch (Exception e) {
                        e.printStackTrace();
                    }
                }

            });
```

## 4. RESULT ANALYSIS AND DISCUSSION

The tests by XDoser are conducted on a website we made just for the purpose. This is used in the test and development stage to understand the workings of the tool. Continuous flood attack was crucial in the process, however increasing the threads might not always work to our advantage. The attack strength is also a variable in the work. As efficient or fast as the method can be, there will be little work if the server can counteract the requests. So, the tool can be used with multiple computers as well. The combined strength of the processing power or many computers will make the attack more efficient.

The tool has been improved over time and now can be used with significant results. Our demo website was used in all the test cases. The tool was run for a specific amount of time and compares with another DoS tool found online. With time being the variable, the tools were tested respectively. The tool has proven to beat the other tools in every case. Spreadsheets were made that were used to make graphs for better representation of the improvements made to the tool over time. The results show that XDoser can provide more loads compares to other tools in a certain given time. Table 3 and Table 4 provides comprehensive data of Table 1 and Table 2 to summarize that XDoser is the 3$^{rd}$ among the tools to in sending more requests than other tools and that it is the top among creating more synthetic load for testing.

The latest results show that the demo website failed about 81.63%, a ratio of 4.44 compared to other tools lagging below. Our results came out to be above 80%, almost every time. Only about 1 in 23 trials came 79.14% with the time frame being constant among all the trials. Higher the ratio, more the server fails the request.

This result shows consistently that the DoS tool is functional and works better than other tools. The result concludes with the fact that the tool needs more development as it faces problem holding on to the port being used and that takes restarting of the computer to fix. Other than that, the tool can be run under a specific amount of time. As time increased, the tool did not crash, but the rate of attack reduces significantly. However, it is to be noted that rate aside, the effectiveness of the tool remains intact as it still produces over 80% results. The table below represents:

| Name | Time | Success | Failure | Ratio |
|---|---|---|---|---|
| XDoser | 25 seconds | 1870 | 8311 | 4.44 |
| Hulk | 25 seconds | 1856 | 7646 | 4.11 |
| Slowloris | 25 seconds | 1898 | 7902 | 4.16 |
| LOIC | 25 seconds | 1888 | 8315 | 4.40 |
| XOIC | 25 seconds | 1887 | 8319 | 4.41 |
| Tor's Hammer | 25 seconds | 1857 | 8102 | 4.36 |
| PyLoris | 25 seconds | 1901 | 8312 | 4.37 |

Table 1: Failure to Success ratio of XDoser and some DOS tools

| Name | Time | Success | Failure | Total |
|---|---|---|---|---|
| XDoser | 25 seconds | 1870 | 8311 | 10181 |
| Hulk | 25 seconds | 1856 | 7646 | 9502 |
| Slowloris | 25 seconds | 1898 | 7902 | 9544 |
| LOIC | 25 seconds | 1888 | 8315 | 9790 |
| XOIC | 25 seconds | 1887 | 8319 | 10206 |
| Tor's Hammer | 25 seconds | 1857 | 8102 | 9959 |
| PyLoris | 25 seconds | 1901 | 8312 | 10213 |

Table 2: Total number of requests to the server

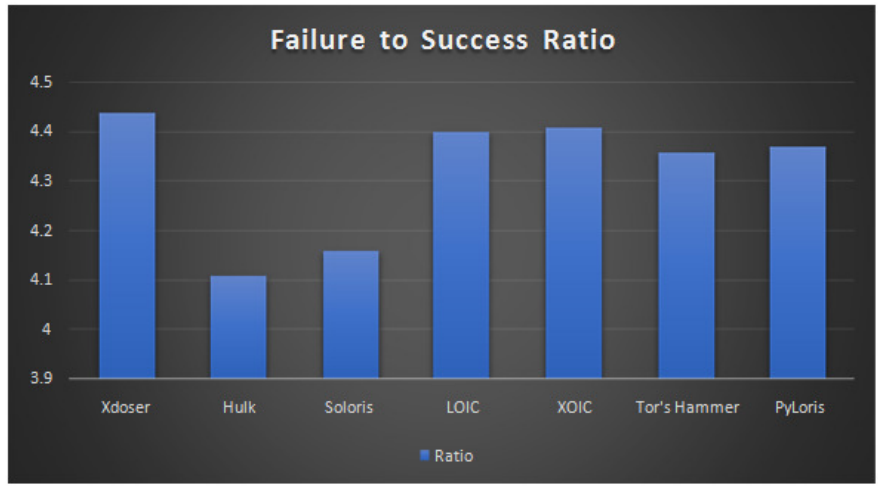

Table 3: Ratio comparing XDoser vs other DOS tools

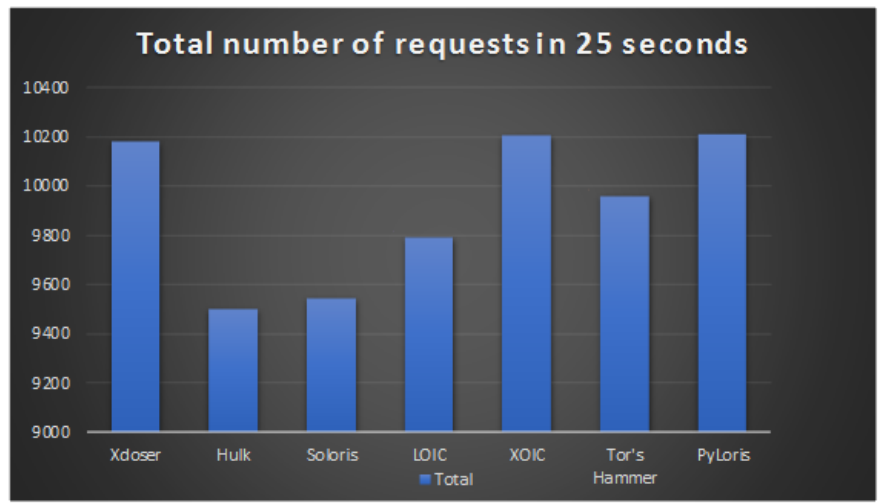

Table 4: Total number of requests to the server

The main limitation of the test was using our basic demo website and a server. The tests were done on a small scale under very basic assumptions that all the requests will be delivered and processed by the server and the tools integrity and endurance. The test field was small and tested for a long period of time and implicates that the load can be consistent, not decreasing as it performs. It is also assumed that the server does not have any basic countermeasure, as that can make a difference. However, that would affect all the tools in the same way. A secure communication scheme has been designed which facilitates communication among corporate and defense agencies [17]. If such a system is not tested for load measurements, it can also be a failure. The architecture is shown below:

## 4.1 COMPARATIVE ANALYSIS

Load Testing or software testing can be done using several tools. In a recent paper "A Pragmatic Evaluation of Stress and Performance Testing Technologies for Web-Based Applications", it shows that any application can stop working as well as cause the failure of the applications without proper testing. It uses HULK script for producing a tremendous load to the web application for testing. HULK stands for HTTP Unbearable Load King. It is a type of DDoS

Attack on the server. Interestingly, it can be used for testing purposes as well. HULK allows checking the performance of the application to see whether the web application is powerful enough to manage the DDoS based traffic. If compared with different tools, HULK is quite faster in testing load compare to Apache JMeter [18]. HULK accounts for 1.82 seconds on average, whereas Apache JMeter accounts for 2.12 seconds. Thus, HULK is better than Apache JMeter. XDoser works like the similar way as HULK because both are DDoS tool. As discussed in result analysis, compare to Hulk and Soloris, the percentage rate of load testing is more in Xdoser which is around 80%.

Xdoser works on HTTP POST requests as the user is continuously sending HTTP POST request for searching a wrong data. Since repetitive searching is going on, so XDoser is creating load concurrently on the backend. In a recent paper "svLoad: An Automated Test-Driven Architecture forLoad Testing in Cloud Systems" [19], it shows that HTTP and HTTPS requests response time is quite more compare to other methodology used for testing. It present comparison of HTTP and HTTPS requests response time. Here, HTTP type test cases are up to 90% faster than HTTPS protocol type. So, XDoser's methodology is quite faster than other methodologies because it uses HTTP POST requests.

In a recent paper, "Efficient DDoS Flood Attack Detection using Dynamic Thresholding on Flow-Based Network Traffic", it states that DoS attacks are one-to-one, that uses one compromised host and influences a network with a small bandwidth. The DDoS attack uses many compromised hosts that will flood the system or network with very large traffic. DDoS has many-to-one attack dimensions, so it is more successful in blocking the victim against its defenses. Botnets are nowadays used to attack in large scale flooding attacks. The attack flows become distributed and more harmful by making it hard to detect. This paper analyses the features of traffic in the network and the existing algorithms to detect DDoS attacks and proposes an efficient statistical approach to detect the attacks based on traffic features and dynamic threshold detection algorithm. The proposed algorithm extracts different features regarding traffic, it calculates four attributes grounded on the characteristics of DDoS and the attack get detected when the calculated attributes within a time interval are greater than the threshold value. In this work, a virtual network using the Virtualbox tool is set up. It uses TCP, UDP, ICMP and RAW-IP protocols. This tool is a platform independent and used as a network packet producer which initiates DDoS, DoS, and MITM.

**4.2 BENEFITS OF USING XDOSER**

DDoS is a viral attack. Any people in the world are affected or victimize by this attack. In general, DDoS tools are used to crash servers. It makes all the resources unavailable and slows down all the current activities. Ultimately, it crashes down the server and harms other people's computers, applications as well as websites. Every coin has two sides The attackers used the DDos tool for the wrong usage whereas XDoser tests the load that an application can bear. It will create a massive load on the website to check the highest capability the application can bearSuppose, an application is affected by any DDoS attack or malware. If the developer of that application knows the highest amount of load the application can sustain, he would have made his application stronger. XDoser will let those developers know about the application's highest load taking capability. Though it is illegal to perform dos operation on a system, XDoser can be used by government or concerned security services for targeting potential harmful systems[20].

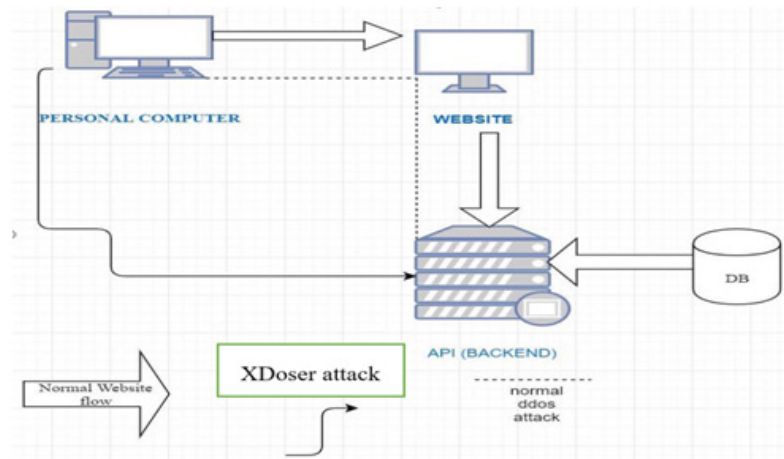

Figure 11. System architecture of XDoser

## 5. CONCLUSIONS

Regardless of the DoS tool having a negative connotation to it, the application of such a program is not something to be ignored. If used properly, it can clean up the internet to some degree. However, the main use of our tool is to test servers and their integrity. Attacks on servers are imminent and will not stop any day soon. This tool will be used to test the load a server can take, and what will be needed to improve it. As it is more robust than other tools found, it can provide better results. Later on, the companies can set their servers and prepare them to their specifications.

## ACKNOWLEDGEMENTS

The authors would like to thank Cyber Security Research Team, North South University, Bangladesh.